# Temperature dependent temporal coherence of metallic-nanoparticle-induced single-photon emitters in a WSe$_2$ monolayer

Martin von Helversen[1], Lara Greten[2], Imad Limame[1], Chin-Wen Shih[1], Paul Schlaugat[1], Carlos Antón-Solanas[3,*], Christian Schneider[3], Bárbara Rosa[1], Andreas Knorr[2], and Stephan Reitzenstein[1]

[1]*Institute of Solid State Physics, Technische Universität Berlin, 10623 Berlin, Germany*

[2]*Institute of Theoretical Physics, Technische Universität Berlin, 10623 Berlin, Germany*

[3]*Institute for Physics, Carl von Ossietzky Universität Oldenburg, 26129 Oldenburg, Germany*

*\*Present address: Depto. de Física de Materiales, Instituto Nicolás Cabrera, Instituto de Física de la Materia Condensada, Universidad Autónoma de Madrid, 28049 Madrid, Spain*

**Abstract**:

In recent years, much research has been undertaken to investigate the suitability of two-dimensional materials to act as single-photon sources with high optical and quantum optical quality. Amongst them, transition-metal dichalcogenides, especially WSe$_2$, have been one of the subjects of intensive studies. Yet, their single-photon purity and photon indistinguishability, remain the most significant challenges to compete with mature semiconducting systems such as self-assembled InGaAs quantum dots. In this work, we explore the emission properties of quantum emitters in a WSe$_2$ monolayer which are induced by metallic nanoparticles. Under quasi-resonant pulsed excitation, we verify clean single-photon emission with a g$^{(2)}$(0) = 0.036 ± 0.004. Furthermore, we determine its temperature dependent coherence time via Michelson interferometry, where a value of (13.5 ± 1.0) ps is extracted for the zero-phonon line (ZPL) at 4 K, which reduces to (9 ± 2) ps at 8 K. Associated time-resolved photoluminescence experiments reveal a decrease of the decay time from (2.4 ± 0.1) ns to (0.42 ± 0.05) ns. This change in decay time is explained by a model which considers a Förster-type resonant energy transfer process which yields a strong temperature induced energy loss from the SPE to the nearby Ag nanoparticle.

## 1. Introduction

Semiconducting materials are not only an integral part of the technological world, they have also proven to host excellent non-classical light emitters. In order to enable applications in photonic quantum technologies related single-photon sources have to feature simultaneously are high brightness, strong multi-photon suppression, and high photon indistinguishability. Over the past decades, it has been well established that semiconductor quantum dots (QDs) can simultaneously fulfil to a large extend the prerequisites mentioned above [1–4]. However, the fabrication of such sources, based on traditional semiconductor materials, requires expensive semiconductor nanotechnology including for instance epitaxial growth and clean room processing. To produce quantum light sources in a simple and cost efficient manner, significant research activities have been directed in recent years towards alternative material and device concepts based for instance on two-dimensional (2D) materials [5].

Among the emerging 2D materials, hexagonal boron nitride (h-BN) and transition-metal dichalcogenides (TMDC) are considered as promising platforms to host high-quality single-photon emitters (SPEs). In fact, many studies have been performed in recent years to understand and improve their optical and quantum-optical properties [6–13], to explore their scalability [14, 15] or to integrate them into nanophotonic structures, such as waveguides [16], circular Bragg grating resonators [17] or Fabry-Perot cavities [18]. Moreover, addressing application relevant temperature ranges, multi-photon suppression was observed up to 150 K [19] for $WSe_2$ monolayers and even at room temperature for h-BN multilayers [20, 21]. Beyond single-photon emission from solid-state sources, photon indistinguishability plays a crucial role in advanced quantum photonic applications such as quantum repeater networks relying on entanglement distribution via Bell-state measurements [22, 23]. While first experiments have verified the fundamental possibility to achieve quantum interference from emitters in atomically thin 2D materials [18] and multilayer hBN [24] in Hong-Ou-Mandel (HOM) measurements, the observation of highest degree of photon indistinguishability remains a challenge and the maximum achieved visibility of 56 % lacks far behind the performance of semiconductor QD sources in cavities. $WSe_2$ is a particular interesting material to host high-quality SPEs [6–11, 15, 17, 25–30], and towards achieving high photon indistinguishability it is important to explore and understand the decoherence mechanisms affecting the performance of SPEs in this material.

Here we explore SPEs in a $WSe_2$ monolayer focusing on the temporal coherence of their emission to obtain a better understanding of the underlying decoherence sources and the current limitations to address. The used sample was fabricated by transferring a $WSe_2$ monolayer onto Ag nanoparticles covered by a few nanometers of $Al_2O_3$, which is known to generate SPEs of high quality in 2D materials [7]. Experimentally, we first verify single-photon emission and then perform Michelson-interferometry to measure the coherence time. Quasi-resonant excitation and triggered pulsed excitation are used to improve the brightness and purity of the studied SPEs, which yield a $g^{(2)}(0)$ as low as $g^{(2)}(0) = 0.036 \pm 0.004$. Our metal substrate provides a remarkable temperature-dependent interaction yielding an SPE lifetime decay varying 10 times from 4 K to 10 K, along with a strong quenching of the zero-phonon line (ZPL). In addition, the temperature-dependence of the coherence was extracted through Michelson interferometer measurements. To obtain a broader understanding, we perform complementary measurements in WSe$_2$-based SPEs observed on a dielectric substrate. Lastly, to support our experimental finding we established a model based on Förster-type resonant energy transfer to explain the temperature dependent properties. Comparing experimental and theoretical results helps us to unveil the semiconducting emitter-metal interaction and, thus, to potentially improve the capability of 2D materials as a suitable source for quantum photonic applications in future.

**2. Experimental results**

2.1 Fabrication of strained WSe$_2$ monolayers

To obtain the target substrate onto which the WSe$_2$ monolayers (ML) are transferred, a (200 ± 10) nm thick silver film was deposited by electron-beam evaporation on a sapphire substrate. The surface was next polished by Ar-ion milling (detailed information can be found in Ref. [7]) leading to the formation of nano- and micro-islands of silver. Lastly, a nominal 3 nm thick Al$_2$O$_3$ layer was deposited to prevent later oxidation as well as to inhibit quenching of the photoluminescence that occurs when the carriers generated in the monolayer are transferred into the metal before the radiative recombination of electron-hole pairs takes place. WSe$_2$ monolayers were produced by mechanical exfoliation of a commercial WSe$_2$ crystal (HQ Graphene) and transferred onto the substrate by performing dry-transfer method [31] using a polydimethylsiloxane (PDMS) stamp. Figure 1(a) shows an optical microscope image of the investigated flake and in panel (b) a scanning electron microscopy (SEM) image reveals the aforementioned nano-islands of silver (dark dots), which are folding the monolayer and therefore induce sections of localized strain.

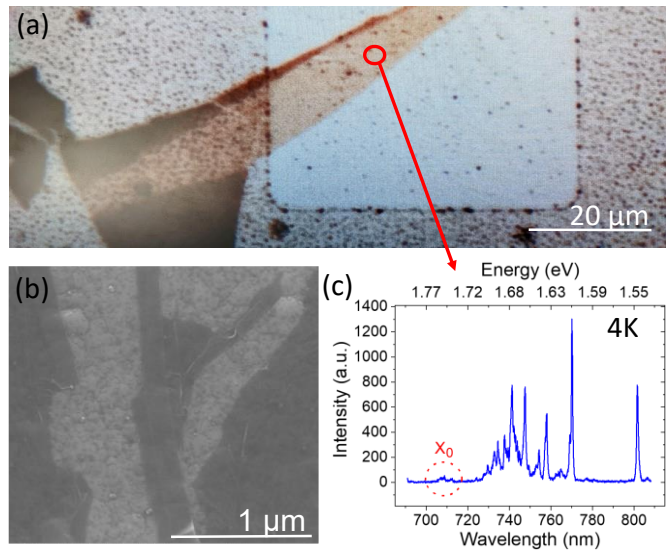

*Figure 1 (a): optical microscope image of the investigated WSe$_2$-monolayer. (b): SEM of a monolayer on the same substrate material allowing to see the silver nano-islands. (c): exemplary photoluminescence at spectrum at 4 K of the investigated flake – the neutral exciton of WSe$_2$ is indicated in red, while bright localized emission arises at lower energies.*

2.2 Optical setup

The sample is then mounted onto a piezo-stack for x-y-z nanopositioning in a closed-cycle cryostat reaching a minimal temperature of 3.9 K. The cold-finger temperature can be temperature stabilized to within a few mK using a proportional–integral–derivative (PID) controlled heater. A confocal microscope with a numerical aperture (NA) of 0.81 allows for optical excitation and detection. Optical excitation is performed by three different lasers: a) a 660 nm (1.88 eV) continuous wave (CW) diode-laser, b) 2 ps pulses from a doped fiber laser that are converted by an optical parametric oscillator (OPO) (up to 1.77 eV) or c) a tunable Ti:sapphire laser in CW operation (up to 1.75 eV). The micro-photoluminescence (µPL) is recorded by a charged-coupled device (CCD) in a spectrometer with down to 40 µeV resolution at 780 nm. This spectrometer can also be used for narrow spectral selection, while bulk optic bandpass filters are used for selecting broader spectral ranges (minimum spectral window: ~1 nm). The signal can also be fiber-coupled to a single-mode fiber, either solely serving as a spatial filter where an area of about 1.5 µm² on the sample surface is imaged onto the fiber core, or as well after spectral filtering using the exit slit of the monochromator. In Figure 1(c) an exemplary micro-photoluminescence (µPL) spectrum taken at 4 K under above-band excitation is shown, stemming from the position indicated by the red circle in panel (a). One can observe the expected emission from the neutral exciton ($X_0$) of the monolayer at a wavelength of 708 nm (1.75 eV), along with several sharp lines at lower energies, commonly attributed to intrinsic SPEs in $WSe_2$. These sharp features are found throughout the monolayer, are localized, and are assumed to stem from the strain induced by the silver nanoparticles.

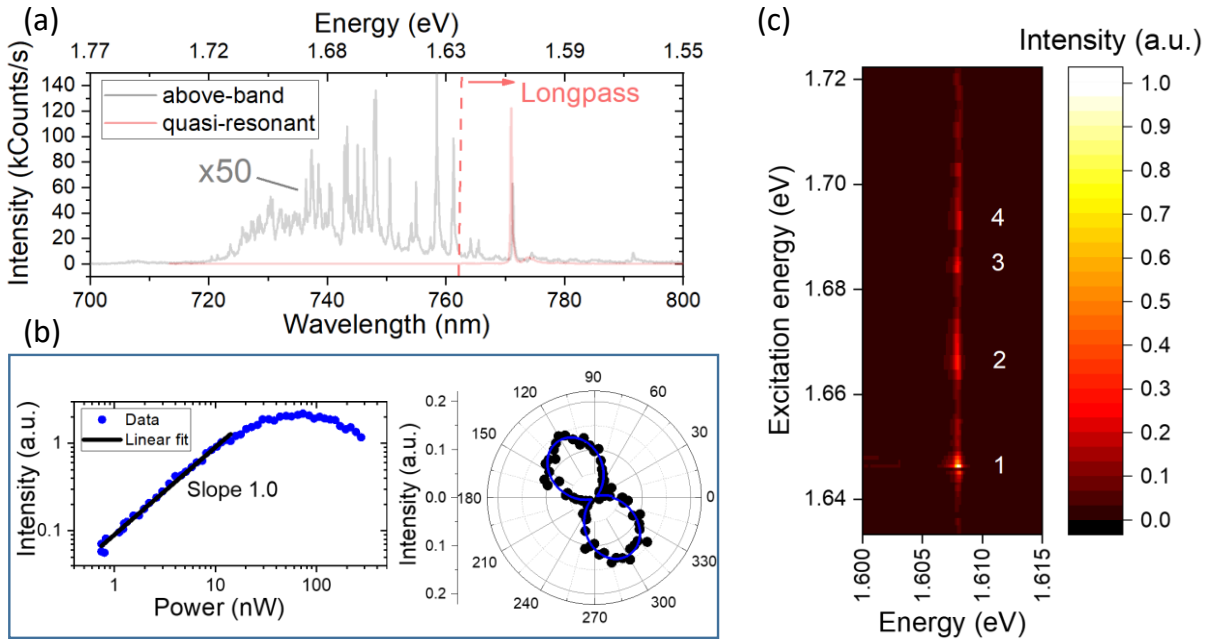

*Figure 2 (a): µPL spectrum of selected emitter under above-band (black, multiplied by 50) and quasi-resonant (red) excitation. (b): Results of power- and polarization- dependent characterization. (c) Color-coded intensity map of a PLE-scan with indication of the first four found resonances.*

2.3 Quantum optical characterization of one selected emitter

*Quasi-resonant excitation*

We performed additional µPL measurements to then identify a suitable quantum emitter for detailed optical and quantum optical studies, and to evaluate its general suitability for quantum optical protocols.

Figure 2(a) shows the µPL spectrum of the selected emitter under above-band excitation (660 nm). This emitter generates photons at 772 nm (1.606 eV) and fitting a sum of a Lorentzian peak for its ZPL and a Gaussian representing the phonon side band (PSB) we extract a linewidth of just above 90 µeV of the ZPL at 4 K. (cf. section *Temperature dependent results*). The excitation power dependence in double logarithmic scaling yields a perfectly linear slope of $1.00 \pm 0.05$ for the emission line at 772 nm, as depicted in Figure 2(b). Also, visualized in the same panel, a degree of linear polarization of $(91 \pm 6)$ % is determined via polarization-resolved µPL measurements. Similar power and polarization dependent properties are found for most of the SPEs on the sample, where two axes of linear polarization appear throughout the monolayer, almost orthogonal to each other. In Ref. [7] a different monolayer on the same type of substrate was investigated and consistent with our results an angle of ~80° was extracted for linearly polarized SPEs. Interestingly, the selected emitter, as well as most SPEs found throughout the monolayer, does not feature a second emission component of orthogonal polarization. We attribute this behavior to the thermal depletion of the second transition in these particular emitters (cf. [32]).

Further insight into the SPE properties is obtained by photoluminescence excitation (PLE) spectroscopy. For this purpose, we wavelength-scanned the CW Ti:sapphire laser, stabilized in power as well as polarization, and simultaneously observed the resulting PL of our selected emitter at 772 nm. The laser wavelength was scanned from 720 nm to 760 nm (1.72 eV to 1.63 eV) and was suppressed in detection using a long-pass optical filter, as indicated in Figure 2(a). A color-coded PLE intensity map is shown in Figure 2(c). Consistent with previous results reported in [25–29] we observe multiple resonances, while the intensity of their response increases with decreasing energetic separation to the emission peak under study. At the closest resonance (labeled 1) which is located at 1.645 eV, we record a splitting of ~40 meV (19 nm) and, in comparison to the off-resonant excitation method we observe a > 50-fold increase of intensity at saturation power $P_{sat}$, showing the efficiency of this quasi-resonant excitation scheme.

*Single-photon characteristics*

One of the most important properties of quantum emitters and simultaneously of particular interest for many quantum photonic applications is the degree of purity in terms of the $g^{(2)}(0)$ value. This figure of merit is determined by photon-autocorrelation measurements using a Hanbury Brown and Twiss (HBT)-type setup, where the emitted photons are sent to a 50:50 beam splitter, whose two output signals are detected by superconducting nanowire single photon detectors (SNSPD), with a time resolution of ≈50 ps per channel. The HBT experiment was performed using the application-relevant pulsed excitation [9, 33, 34], where the wavelength of the OPO was tuned to the quasi-resonance (peak 1 in Figure 2(c)). The emission of the emitter was spectrally filtered by the exit slit of our monochromator (bandwidth set to 0.2 nm, i.e. three times the linewidth of the emitter) and the correlation histogram shown in Figure 3 was obtained. The strongly reduced coincidences at zero time-delay reflect the high single-photon purity of the quantum emitter, which is further underlined by the semi-logarithmic scaling

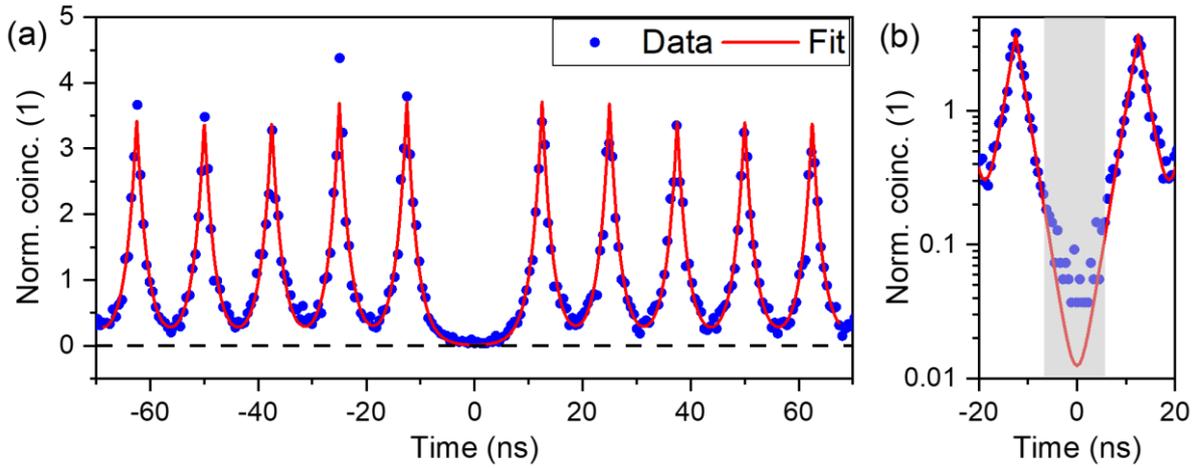

*Figure 3 (a): Second-order autocorrelation function under pulsed quasi-resonant excitation revealing an antibunching value of $g^{(2)}(0) = 0.036 \pm 0.004$. (b): Central correlation coincidences around zero-time delay in semi-logarithmic scaling – the counts surpassing the fit at zero delay are assigned to stem from multi-photon events.*

of the two-photon coincidences in Figure 3(b). To quantify the multi-photon suppression, we fit the ten adjacent peaks using a double-sided exponential function. For the zero-delay correlation coincidences, all counts that exceed the fitted function in a time interval of $(-\Delta t, +\Delta t)$ are attributed to multi-photon emission. We do this for the full time-window ($2\Delta t = 12.5$ ns), corresponding to the 80 MHz repetition frequency of the driving laser, as well as for post-selected time windows of 10 (8) ns centered around zero time-delay. The full window of 12.5 ns yields a value of $g^{(2)}(0) = 0.036 \pm 0.004$, while the chosen degree of temporal post-selection goes down to $g^{(2)}(0)_{10\ ns} = 0.027 \pm 0.005$ ($g^{(2)}(0)_{8\ ns} = 0.022 \pm 0.006$), proving the excellent quantum nature of emission that can meanwhile be achieved using TMDC based single-photon sources (Table 1 in the supplementary material shows the current record values of single-photon purity). It is worth mentioning that these very promising results heavily depend on the excitation conditions – using above-band excitation the measured autocorrelation function appears to be less clean as re-excitation processes lead to multiple emission per excitation cycle (cf. [7]).

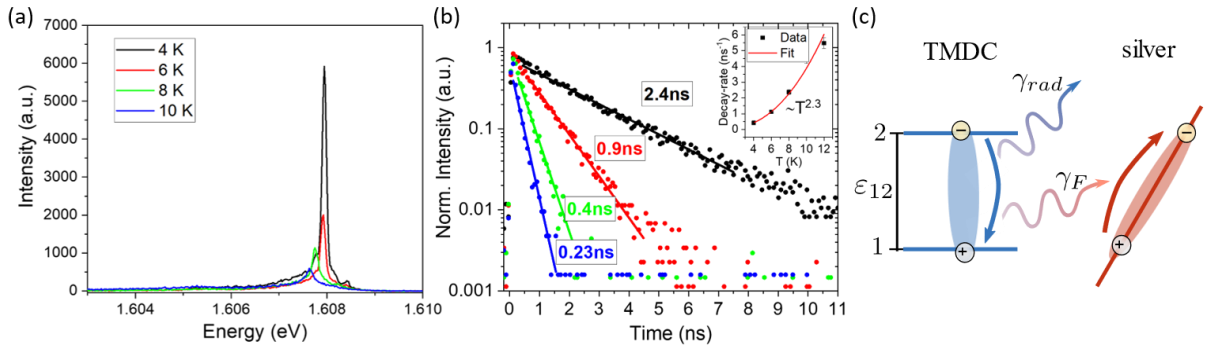

*Figure 4 (a): Temperature-dependent spectra under quasi-resonant excitation. (b): Corresponding measurements of the emitters decay times $T_1$. The inset depicts the decay rates over temperature and a fit ($y = const.+AT^n$) from which the power n is extracted to be $2.3 \pm 0.3$. (c): Sketch of decay processes for the emitters' occupation $\sigma_{22}$ in the TMDC. Its excitation can decay radiatively or is transferred to the silver substrate by a Förster-process where it induces an intra-band transition.*

*Temperature dependent results*

Non-classical light emitters yielding high multi-photon suppression are induced by the silver particles below the TMDC monolayer [7]. However, in addition to this wanted effect, the metal can also have detrimental influence on the emission efficiency and on the coherence of the SPEs. To explore this aspect, we first performed temperature dependent μPL studies. Corresponding μPL spectra of the selected emitter (obtained under quasi-resonant excitation) are plotted in Figure 4(a) for T = 4, 6, 8, 10 K. Interestingly, we observe about an order of magnitude decrease in μPL intensity with only this small change of temperature. Moreover, the intensity ratio of ZPL/PSB strongly decreases from about 1.5 at 4 K (cf. Ref. [6] and supplemental material for the fit to the data) to 0.6 at 10 K, and above 14 K almost all emission of the ZPL is suppressed. To explain this effect and to investigate the temperature dependent emission properties further, we next studied the time-resolved response, using the same excitation conditions as for the second-order autocorrelation discussed above. Here, as depicted in Figure 4(b) a very fast, temperature-induced decay of the emitter's lifetime becomes apparent.

To explain this temperature-dependency, we provide a minimal optical Bloch equation model for the emitter-metal hybrid structure: The emitter's upper occupation density $\sigma_{22}$ after a preceding excitation is given by

$$i\hbar \partial_t \sigma_{22} = -2i\big(\gamma_{rad} + \gamma_F(T)\big)\sigma_{22}\,, \tag{1}$$

which derivation is analog to Ref. [35]. The corresponding energy and time dependent luminescence can be found in Ref. [36]

$$\mu\text{PL}(\varepsilon,t) \propto Im\left(\frac{1}{\varepsilon_{12} - \varepsilon - i\big(\gamma_{rad} + \gamma_F(T) + \gamma_{pure}\big)}\right)\sigma_{22}(t)\,, \tag{2}$$

where $\varepsilon_{12}$ is the transition energy of the selected emitter and $\varepsilon$ the energy of an emitted photon. The decay of the emitter's upper occupation $\sigma_{22}$ governs the lifetime $T_1$ and mirrors the time dependency of the μPL, displayed in Figure 4(b), whereas its linewidth in Figure 4(a) is given by the energy dependent prefactor of the μPL and corresponds to the coherence time $T_2$. The decay of $\sigma_{22}$ consists of two contributions: the direct radiative decay $\gamma_{rad}$, that would appear for any substrate and an additional decay channel $\gamma_F(T)$. The latter stands for a Förster process [37], where the emitter's excitation is non-radiatively passed on to the silver substrate, mediated by the electric near-field [35, 38]. The corresponding energy transfer process is sketched in Figure 4(c). For this process to occur, a non-vanishing density of unoccupied states with matching excitation energy in the substrate is required. For the silver substrate, this corresponds to an intra-band transition within the silver conduction band [39, 40]. The observed temperature dependence of the decay in the silver substrate via electron-phonon scattering $\gamma_{silver}(T) \propto T^2$ [41] is therefore transferred to the Förster decay channel of the emitter's excitations. It follows effectively $\gamma_F(T) \propto T^2$, which is in very good agreement with the measurements provided in Figure 4(b) (see inset of Figure 4(b)).

Further, to support our interpretation that the temperature dependency occurs due to the energy transfer between the emitter and the metal mediated by Förster interaction, in a second sample we place a monolayer of the same WSe$_2$ crystal on top of a dielectric substrate patterned with nanopillars for comparison. The results of which are displayed in the supplementary material, showing no significant change of the emitter's lifetime within the same temperature range. This comparative study supports our explanation of the temperature dependence in Figure 4(b) since the dielectric substrate provides no possible transition with suitable excitation energy, which could allow for the temperature dependent decay channel. Comparing Figure 4(a) and (b) shows that the ideal relation $T_2 = 2T_1$ between coherence time $T_2$ (reflected in the linewidth) and lifetime $T_1$, achieved for high-quality QDs at low temperatures [42, 43] is by far not reached in our sample. This is related to e. g. phonon-induced pure dephasing $\gamma_{pure}$ [44] and observed consistently for SPEs in 2D materials [45]. Noteworthy, due to the significantly larger magnitude of pure dephasing $\gamma_{pure}$ compared to the radiative $\gamma_{rad}$ one and Förster rates $\gamma_F$, their respective contributions are negligible in equation (2).

As mentioned beforehand, a key requirement in quantum photonic applications is the photon indistinguishability. Instead of solely calculating the coherence time from the extracted linewidths, we additionally access this optical property via free-space Michelson-interferometry, as shown in Figure 5(a). For stable alignment and to simultaneously select only one spatial mode, the interferometer´s input is based on a single-mode fiber. Both arms are equipped with retroreflectors. The

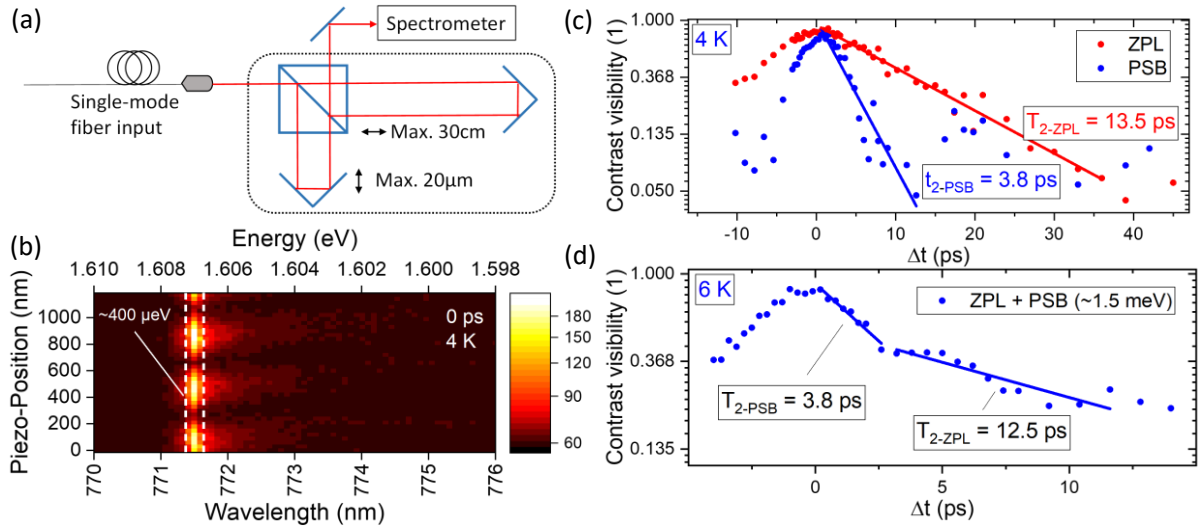

*Figure 5 (a): Sketch of the free-space Michelson-interferometer based on two retroreflectors and a 50:50 beamsplitter. (b): color-coded intensity map of an interferometer scan at 4 K with zero path difference. The white dashed lines indicate an exemplary region of a spectral integration window of 400 µeV. (c): Contrast visibility at 4 K for the ZPL and the PSB separately, each using a spectral window of about 350 µeV. The solid lines show linear fits to the data and the extracted $T_2$-times. (d): Contrast visibility at 6 K for a spectral window of 1.5 meV containing both the ZPL and the PSB. The bi-exponential decay constants are again obtained using linear fits (solid lines).*

path-length difference is set using a linear motorized stage (with tens of µm precision), while a piezo-element with nm precision enables a phase-scan over a few periods of the wavelength of the investigated line. One of the two outputs of the apparatus is then fed into our spectrometer where the data is recorded by a CCD. In Figure 5(b), we plot such an interference map, taken at 4 K at equal arm lengths. By acquiring the whole spectral information, we are now able to select a region of interest (exemplarily indicated with white dashed lines in the same panel, here centered around the ZPL), spectrally integrate in this interval and then calculate the corresponding interference contrast via the normalized visibility $C = (I_{max} - I_{min}) / (I_{max} + I_{min})$, which is done from a sinusoidal fit to the obtained data (relative phase change between the Michelson arms, fine delay variation) for each path difference (coarse delay variation $\Delta t$). The maximum contrast visibility we achieve is ~ 96-97 %, also confirmed in a reference measurement with a MHz-linewidth diode-laser. We attribute this non-ideal visibility to alignment imperfections as well as the polarization dependence of the retroreflectors. To evaluate the coherence time $T_2$ of the selected SPE and compare the results to the spectral information obtained above, we took scans at 4, 6 and 8 K, again using CW quasi-resonant excitation. The advantage of our method is shown in the following: In Figure 5(c) two different spectral regions are selected, namely around the ZPL (red data points) and at slightly lower energy in the PSB (blue data points), each filtered with a spectral width of about 350 µeV. The contrast visibility is plotted logarithmically, and we fit linear slopes to the exponential decays, which results in $T_{2\text{-PSB-4K}} = (3.8 \pm 0.4)$ ps and $T_{2\text{-ZPL-4K}} = (13.5 \pm 1.0)$ ps. The latter is in very good agreement with the calculated coherence time of 14.3 ps extracted from the linewidth (92 µeV) of the Lorentzian component of the emitter´s line shape. But also, one can choose an integration window containing both the ZPL and the PSB simultaneously, as is shown in Figure 5(d) for the measurement at 6 K, where a spectral range of 1.5 meV was selected. Again, plotting the contrast in logarithmic scaling, the two different exponential decays become clearly visible and allow for extracting two time-constants, which we attribute again to ZPL and PSB. The numbers are $T_{2\text{-PSB-6K}} = (3.8 \pm 0.3)$ ps and $T_{2\text{-ZPL-6K}} = (12.5 \pm 1.1)$ ps, which are also at this temperature in very close agreement to the spectral data, yielding a $T_2$ of just below 12 ps for a linewidth of 110 µeV. The last temperature step to 8 K results in $T_{2\text{-ZPL-8K}}$-values of $(7.5 \pm 1.3)$ ps from the spectra and $(9 \pm 2)$ ps from the Michelson-interferometer.

These temperature-dependent studies reveal an interesting aspect concerning the aforementioned ratio of $(T_2/2T_1)$, where 100 % would describe a Fourier transform limited emitter and 0% would describe a single photon wavepacket unable to interfere. For our emitter the ratio is increased from $(0.28 \pm 0.02)$ % at 4 K to $(0.70 \pm 0.06)$ % at 6 K and $(1.1 \pm 0.2)$ % at 8 K for the ZPL. In fact, due to the above-described quenching, the emission intensity decreases with temperature, but depending on the exact properties of an emitter, the lowest temperature does not necessarily yield the highest $T_2/2T_1$-ratio. This knowledge can be beneficial for future attempts to address the decoherence sources in TMDCs SPEs, a condition for the obtention of high photon indistinguishability.

## 3. Conclusion

We have identified and optically and quantum-optically characterized SPEs in a WSe$_2$ monolayer. To our knowledge they provide the best so far obtained values of second-order autocorrelation in triggered emission at > 50 MHz repetition rate in TMDC-based single-photon emitters, reaching $g^{(2)}(0) = 0.036 \pm 0.004$ for the whole time interval corresponding to the excitation frequency and an even lower $g^{(2)}(0)_{8\,ns} = 0.022 \pm 0.006$ by temporal post selection of an 8 ns window. Given the state´s lifetime of 2.4 ns, this post-selection includes > 95 % of the expected photons from the emitter. Simultaneously, by direct access to the coherence time $T_2$ via Michelson-interferometry and comparing the obtained values to the acquired spectral information we observe values up to about 14 ps at 4 K, which is the highest value yet reported in the literature of a direct measurement using Michelson-interferometry for WSe2-SPEs. Additionally, our method of acquiring the entire spectral information after the Michelson-interferometer allows for flexible spectral selection of the emission energy range of interest. This enables us to investigate different emission features of the same state or even multiple states at different energies simultaneously. However, in [9] an even smaller linewidth (55 μeV compared to the 92 μeV observed for the ZPL in our SPE) is reported, leading to an expected coherence time of about 24 ps. Comparing the determined coherence time of 14 ps with the measured SPE spontaneous emission lifetime of 2.4 ns yields a very low $(T_2/2T_1)$-ratio below 1% that impedes a high visibility in Hong-Ou-Mandel two-photon-interference beyond the current state of the art [18]. Interestingly, the $(T_2/2T_1)$-ratio of the ZPL increases with temperature from $(0.28 \pm 0.02)$ % at 4 K to $(0.70 \pm 0.06)$ % at 6 K and to $(1.1 \pm 0.2)$ % at 8 K, because of the decrease of temporal coherence is less pronounced than the temperature-induced reduction of the spontaneous emission lifetime. The latter is explained theoretically by a Förster-type energy transfer process from the WSe$_2$-SPEs to the nearby Ag nanoparticle. In order to possibly obtain significant indistinguishability of photons generated by SPEs based on atomically thin 2D materials, it will be important to gain a detailed understanding of processes leading to low coherence times, which are most likely related to charge noise. In this regard, the techniques described in Ref. [46, 47] can be very helpful because they can reveal line-broadening mechanisms at different timescales.

**Acknowledgements**

We acknowledge support from the German Research Foundation (DFG) via Grants RE2974/12-1, RE2974/26-1 and from the project EMPIR JRP 17FUN06 SIQUST (the EMPIR initiative is co-funded by the European Union's Horizon 2020 research and innovation program and the EMPIR Participating States). We acknowledge also financial support within the QuantERA II Programme that has received funding from the European Union's Horizon 2020 research and innovation programme under Grant Agreement No 101017733, and from the German Ministry of Education and Research (BMBF) via the projects EQUAISE and TubLan Q.0. The authors are thankful for technical support by the group of

Tobias Heindel. We further thank Prof. D. S. Kim (UNIST, Ulsan, Republic of Korea) and Dr. L.N. Tripathi (Vellore Institute of Technology, Tamil Nadu, India) for preparing and providing the substrate.

**Data availability statement**

The data that support the findings of this study are available upon reasonable request from the authors.

# Supplementary Material

## 1. Comparison to WSe$_2$ monolayer on a dielectric substrate

To compare our obtained results of single-photon emitters on silver nanoparticles with those of non-metal induced ones, we fabricated a dielectric substrate featuring nanopillars with a diameter of 200 nm and a height of nominally 400 nm. The nanopillars were fabricated using electron beam lithography (EBL). Firstly, the 285 nm SiO$_2$/Si substrate is meticulously cleaned before spin coating a 300 nm thick layer of photoresist (AZnLOF 2070) at 4000 rpm. Subsequently, the coated sample is baked for one minute at 100 °C. Afterwards, EBL is performed to pattern an array of nanopillars, spaced by 4 µm, each structure nominally presenting 200 nm diameter. We used an exposure dose of 35 µC/cm². The development step included post-baking the sample at 110 °C, followed by dipping it in the developer (AZ 726 MIF) for 15 s and finally a cleaning step for five minutes in a cascade of deionized water. The size and periodicity of fabricated nanostructures were lastly confirmed by scanning electron microscopy measurements.

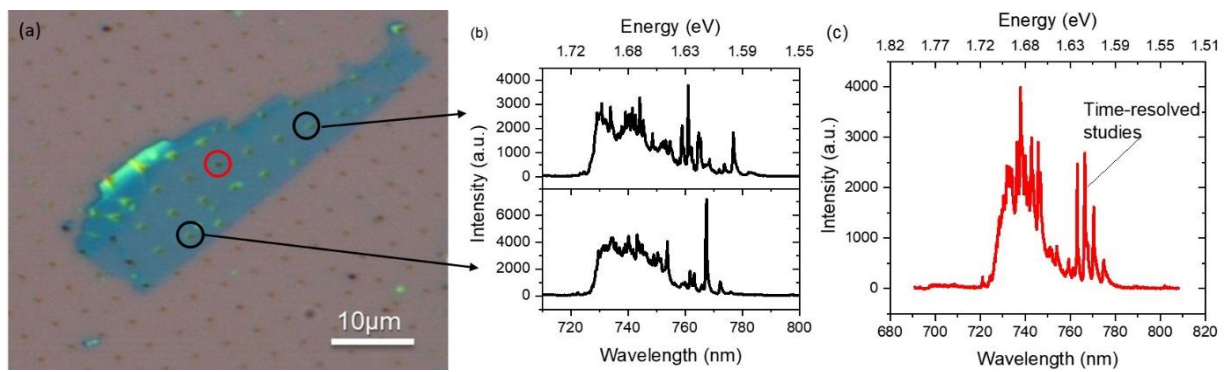

*Figure S1 (a): optical microscopy image of a WSe$_2$ monolayer on top of fabricated pillars. (b),(c): µPL spectra at 4 K under 660 nm CW excitation, taken from the locations marked by black circles (b) and by the red circle (c). The latter emitter is further investigated.*

Next, using the same method as described in the main text, a monolayer of WSe$_2$ was transferred and the sample was investigated in the same optical setup at cryogenic temperatures. An optical microscopy image of the nanopillar sample is shown in Fig. S1(a). We can again find sharp localized emission lines that appear at the locations of the pillars: Fig. S1(b) depicts the 4 K µPL spectra of the locations marked with black circles and Fig. S1(c) corresponds to the red marked site – the emitter from which the data presented in the following was taken. The emitter of interests emits at 1.616 eV (767.2 nm). While this spectrum was acquired under 660 nm above-band excitation, a PLE-scan similar as for the emitter on the silver nanoparticles was performed, too, this time revealing a splitting to the first energetic resonance of ~53 meV. Again, we observe an increase of brightness, however not as pronounced as in the case of the silver nanoparticles. The spectral temperature-dependence of this emitter is shown in Fig. S2(a). For the range of 4 K to 12 K there is a clear broadening of the line from about 350 µeV of the ZPL to up 1 meV for above 10 K. However, the absolute brightness almost stays the same – we actually observe a slight increase of below 10 % going from 4 to 10 K, in stark contrast to the observations of the quantum emitter on metal nanoparticles in the main text. The decay time $T_1$ = 3.0 ns at 4 K is about 50 % longer than $T_1$ of the emitter in the main text. Interestingly, almost no temperature-dependence of the decay time can be observed for the investigated temperature range. The linear fits to the data shown in Fig. S2(b) reveal lifetimes from $T_{1,4K}$ = 3.0 ns to $T_{1,12K}$ = 3.5 ns, highlighting the very different behavior of the two samples.

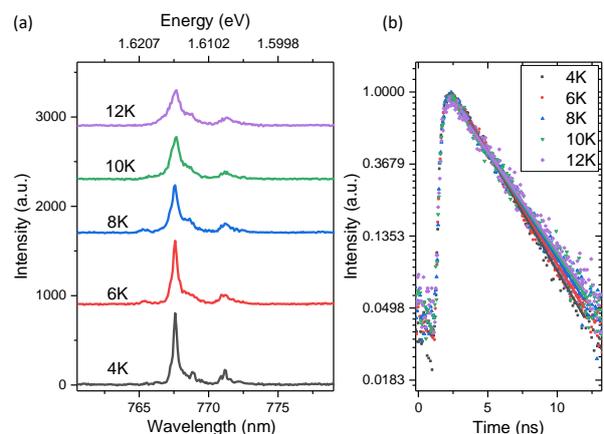

*Figure S2(a): stacked µPL spectra taken under quasi-resonant CW excitation from 4 to 12 K. (b): Time-resolved µPL measurements in the same temperature range.*

Lastly, we also measured the first-order coherence of the emission by using Michelson-interferometry. Using the same approaches as in the main manuscript we extract a very limited coherence time $T_{2\text{-}ZPL}$ of 4 ps from the spectral information (linewidth of 340 µeV at 4 K for the Lorentzian representing the ZPL). This value is confirmed by the interferometric measurement as shown in Fig. S3. Due to the much-decreased coherence time the bi-exponentiality is less prominent, but still a kink in the decay is clearly observable allowing to separate the phonon-contributed emission from the zero-phonon line. The results are perfectly in the expected region and prove again the usability of this spectral method for coherence time evaluation.

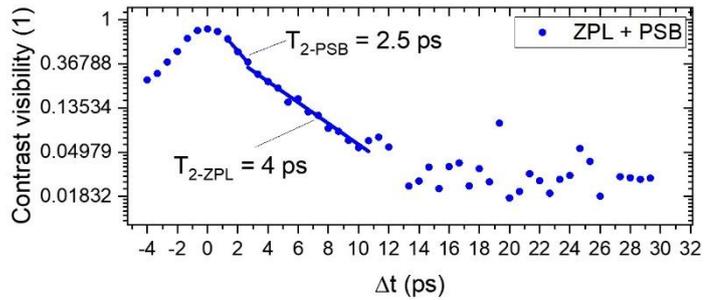

*Figure S3: Contrast visibility of first-order coherence for the selected emitter at 4 K.*

## 2. Fit to the temperature-dependent spectral data

To fitting routine to the temperature-dependent spectra has a Lorentzian component representing the ZPL as well as a Gaussian part to account for the PSB. This method is exemplarily shown in Fig. S4 for a temperature of 4 K. The extracted values are 92 ± 2 µeV for the Lorentzian line width, and just above 150 µeV for the Gaussian representation of the PSB. The ratio of areas ZPL/PSB yields 1.5 at 4 K and decreases to 0.6 at 10 K.

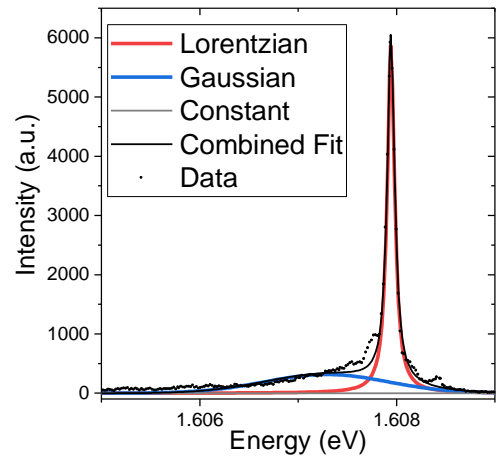

*Figure S4: Fit of the spectral data at 4 K.*

## 3. Literature values achieved in comparison to our stated quantum-optical properties

| Reference | Linewidth (µeV) | $T_1$ (ns) | $T_2$ (ps) | $g^{(2)}(0)$ | Comments |
|---|---|---|---|---|---|
| Koperski et al. 2015 [1] | 100 | 0.6 | | 0.23±0.08 | $T_1$ upper bound from $g^{(2)}$ |
| He et al. 2016 [2] | <70 res. lim. | X: 1.504 XX:0.793 | | X: 0.286±0.04 XX: 0.397±0.06 | Crosscorrelation proving cascaded emission |
| Kumar et al. 2016 [3] | | | | 1) 0.022±0.004 2) <0.002 | High-purity single-photon emission from the ground-state exciton under the resonant excitation of the BS-X. Resonance Fluorescence also achieved. |
| Branny et al. 2017 [4] | 1L:83 | | | 1L: 0.07±0.04 2L: 0.03±0.02 | Mono- and Bilayer On pillars |
| Palacios-Berraquero et al. 2017 [5] | 180 | | | 0.0868±0.0645 | |
| Cai et al. 2018 [6] | 500 | 0.8 | | 0.3 | Plasmonic nanopillars |
| Luo et al. 2018 [7] | 55 | 4.0±1.8 off 0.266±0.12 on | 24 Avg. 14 | 0.16± 0.03 Avg 0.21 | Coherence time, Plasmonic coupling, $T_2/2T_1$ |
| Dass et al. 2019 [8] | | 1: 0.42 2: 2-225 | | 0.13 | Ultralong lifetimes hBN heterostructures |
| Iff et al. 2019 [9] | 200 | 1 | | 0.13 | Piezo wrinkle |
| Sortino et al. 2019 [10] | | 0.2 | | | GaP Nanoantenna |
| Dang et al. 2020 [11] | | | | | g-factor |
| Daveau et al. 2020 [12] | > 1000 | | | 0.08±0.04 | hBN – spectral wandering |
| Iff et al. 2021 [13] | | 0.6 | | < 0.25 | CBGBs + Purcell |
| So et al. 2021 [14] | 400 | | | 0.298±0.036 | AFM indentation, FSS |
| Parto et al. 2021 [15] | 75 | X:6.1 XX:4 | | X:0.05±0.04 XX:0.09±0.05 | Linewidth average, XX, X with FSS Up to 150K: $g^{(2)}$: 0.27 |